\documentstyle[aps,prbbib,psfig]{revtex}

\newcommand{\bvec}[1]{{\mathbf #1}}

\begin{document}
\draft
\widetext
\title{Monte Carlo simulations of an impurity band model for III-V diluted magnetic semiconductors}
\author{Malcolm P. Kennett$^{1,3}$, Mona Berciu$^{2,3}$ and R. N. Bhatt$^{1,2,3}$}

\address{$^1$Physics Department, Princeton University, Princeton, NJ 08544, USA \\
$^2$Department of Electrical Engineering, Princeton University, Princeton, NJ 08544, USA \\
$^3$Princeton Materials Institute, Princeton University, Princeton, NJ 08544, USA}

\date{\today}

\twocolumn[\hsize\textwidth\columnwidth\hsize\csname@twocolumnfalse\endcsname

\maketitle

\begin{abstract}
We report the results of a Monte Carlo study of a model of
(III,Mn)V diluted magnetic semiconductors which uses an impurity band
description of carriers coupled to localized Mn spins and is applicable
for carrier densities below and around the metal-insulator transition.
In agreement with mean field studies, we find a transition to a
ferromagnetic phase at low temperatures.  We compare our results for
the magnetic properties with the mean field approximation, as well as
with experiments, and find favorable qualitative agreement with the
latter.  The local Mn magnetization below the Curie temperature
is found to be spatially inhomogeneous, and strongly correlated with
the local carrier charge density at the Mn sites.
The model contains fermions and classical spins and hence we introduce a 
perturbative Monte Carlo scheme to increase the speed of our simulations.
%which may have wider applicability than the problem considered here.   
\end{abstract}

] 

\narrowtext
\section{Introduction}
\label{sec:intro}

III-V diluted magnetic semiconductors (DMS) have become a very active
area of research due to their interesting magnetic and transport
properties. \cite{Ohno1,Prinz} Thus far Ga$_{1-x}$Mn$_x$As has received the
greatest amount of attention, due to the observation that it becomes ferromagnetic
with a Curie temperature as high as 110K, when $x=0.053$.\cite{highTc,Ohno2}
More recently, ferromagnetism above room temperature has been observed
in (Ga,Mn)N.\cite{nitride,NOhno}

The observed ferromagnetism is widely accepted to be due to a charge
carrier mediated coupling between the Mn spins. Several models have
been proposed to explain the detailed phenomenology of these
compounds.
\cite{Mona1,Mona2,Science,DOM,Konig,Tc,Schliemann,DMFT,SHill,Sanvito2,Semenov,%
Resonance,Perc,frustrate,Noncoll,Chudnovskiy,Kaminski,TSO}
In particular, we have initiated an
effort to understand the effects of disorder in Mn positions on the
properties of these compounds.\cite{Mona1,Mona2,MMR,Comp,Review} An additional source of 
disorder is the large degree of compensation seen in these materials, which
has been attributed to As antisite defects, \cite{SHill} because of which, the carrier density is 
{\it significantly} less than the Mn density.

To this end, we study the low Mn concentration regime, near and below the metal-insulator
transition (MIT) at  $x \simeq 0.035$, where disorder effects would be expected to be the 
most pronounced. In this limit, we model the charge carriers in terms of an impurity band 
comprised of
states around each Mn acceptor, which is taken to be the source of charge carriers
mediating ferromagnetism. Evidence for the relevance of
an impurity band has been provided by a number of recent experiments,
such as an scanning tunneling microscope study \cite{Grandidier}  which showed the 
existence of an impurity band  in (Ga,Mn)As samples with
$x=0.005-0.06$. Angle resolved photoemission spectroscopy in a sample
with $x=0.035$ has also revealed a well formed impurity band, and
confirmed that the Fermi energy lies in its vicinity.\cite{ARPES}

A complete, detailed description of the impurity band in the presence of compensating 
defects is an extremely
difficult enterprise, \cite{comment} hence we have proposed a simple effective Hamiltonian
that captures (at least qualitatively) the relevant impurity band
physics. This Hamiltonian was studied comprehensively at the
mean-field level, and the magnetic properties were observed to have a number of 
surprising properties.\cite{Mona1,Mona2} The mean-field magnetization
curves have very unusual, concave upward shapes, unlike the
magnetization curves of conventional ferromagnets.  Some of these features have been seen in 
experimental measurements, especially for samples with a low carrier density and
high degree of compensation. \cite{Beschoten} 
This unusual
shape of the magnetization curves was identified to be a direct effect
of positional disorder of the Mn ions.\cite{Mona2} 

The mean 
field calculation suggested that there is considerable inhomogeneity in the 
magnetization of individual Mn spins
at temperatures below $T_c$, particularly for small values of $x$.  
Experimentally, disorder appears to be relevant even in the metallic phase --
Barkhausen jumps were observed in a sample with $x = 0.047$, \cite{Bark} indicating
the presence of frozen-in magnetic disorder.   
Theoretically, disorder has also been credited 
as leading to an instability in collinear Mn ground states in the metallic phase,
\cite{Noncoll,Chudnovskiy} via spin waves, which can lead to a reduction in the saturation
magnetization at low temperatures as observed in annealed samples with $x = 0.05$. 
\cite{Potashnik}

Increased disorder was also
shown to lead to an increased value of the critical temperature
$T_c$ in the mean field study.\cite{Mona1} 
It is well known that the mean-field approximation underestimates the effect
of thermal fluctuations and therefore overestimates the value of the
critical temperature.  As a result, it is important to check the mean field
results against Monte Carlo simulations which properly account for
thermal fluctuations, to test to what extent the phenomenology found in the
mean-field study is maintained when these fluctuations are included. 

In this paper, we report the results of such a Monte Carlo study.  We find that the
magnetization curves retain their unusual shape even when thermal fluctuations are
included and that, as expected, the critical temperature is lowered from the mean field
value.  The extent of this decrease is most pronounced for low values of $x$ and high
compensation.  The observation that disorder can lead to a higher critical temperature
than a purely ordered case \cite{Mona1} is also confirmed, although the change is more modest 
than in the mean field study.  The spatial inhomogeneity 
observed in the mean field studies \cite{Mona2}
is confirmed and found to be little changed by thermal fluctuations.  A correlation 
between regions of larger local charge density and larger local magnetization is also
established.

This paper is organized in the following manner:
in section \ref{sec:model} we describe the
model Hamiltonian and the values chosen for various parameters of the
system.  We also discuss the use of finite size scaling to determine the ferromagnetic
transition and describe the quantities that we calculate.  In section \ref{sec:PMC}, we 
introduce a perturbative scheme to speed up the Monte Carlo simulations, 
describe our implementation and discuss 
testing on a toy model for which we can compare with exact results.  
We present the results of our Monte
Carlo simulations for the impurity band model of DMS in sections \ref{sec:results} and
\ref{sec:five}.  Section \ref{sec:conc}  
summarizes our conclusions and discusses 
the implications of our results for experiments and modelling of III-V DMS.

\section{Model}
\label{sec:model}

\subsection{Hamiltonian}
When manganese is introduced into GaAs, the Mn impurities have been
shown to substitute on the Ga FCC sub-lattice of the zinc-blende structure of the undoped
semiconductor for small values of $x$. \cite{Ohno3,Shioda}
However, at larger values of $x$ ($\geq$ 0.07), Mn ions can form MnAs clusters, which have
the NiAs structure. \cite{Ohno3,Shioda,VanEsch}  Based on these experimental findings, 
we assume a zinc-blende 
structure and Mn substituting only on Ga sites for the low Mn concetrations we study.  
Each Mn impurity has a 
spin-$\frac{5}{2}$ from its half-filled 3$d$ shell.  The nominal valence-II of Mn implies 
that when it substitutes for the valence-III Ga, it acts as an acceptor.  Thus an isolated
Mn can bind a hole in an impurity level that we characterize by
a hydrogenic orbital with Bohr radius $a_B$, with wavefunction $\phi_i(\bvec{r}) \sim 
\exp[-|\bvec{r} - \bvec{R}_i|/a_B]$.  In Mn doped GaAs there are substantial
central cell corrections \cite{Guillaume} which we phenomenologically incorporate by 
adjusting the Bohr radius $a_B$.  

Whilst the true carriers in this system are
holes with spin $\frac{3}{2}$, we consider the case of electron doping, so the carrier spin is 
$\frac{1}{2}$.  
This leads to some differences \cite{Schliemann} (in particular, the frustration effects recently
claimed for hole doping \cite{frustrate} are absent in our model).   However, 
the effects of disorder and 
impurity potentials, which are our
main focus here, should not be qualitatively changed. \cite{Mona1,Mona2}  The carriers (electrons)
are then in an impurity band below the conduction band minimum of the host semiconductor.

We study the model introduced by
Berciu and Bhatt, \cite{Mona1,Mona2} for which the Hamiltonian is

\begin{eqnarray}
\label{eq:Ham}
{\mathcal H} & = & \sum_{i,j} t_{ij} c^\dagger_{i\sigma} c_{j\sigma} + \sum_{i,j} J_{ij} 
\bvec{S}_i \cdot \left(c^\dagger_{j\sigma}\frac{1}{2}
\mbox{\boldmath$\sigma$}_{\alpha\beta}c_{j\beta}
\right) \nonumber \\
& & - g^*\mu_B \bvec{H} \cdot \sum_i c^\dagger_{i\sigma}\frac{1}{2}
\mbox{\boldmath$\sigma$}_{\alpha\beta}c_{i\beta} - g\mu_B 
\bvec{H} \cdot 
\sum_i \bvec{S}_i.
\end{eqnarray}
The random positions of the Mn impurities are labelled by $\bvec{R}_i$,
and the spin of the Mn impurity at $\bvec{R}_i$ is $\bvec{S}_i$.  In the electron formalism
for the charge carriers, 
$c^\dagger_{i\sigma}$ is the creation operator for a carrier with spin 
$\mbox{\boldmath$\sigma$}$ in the bound 
state associated with the $i^{th}$ impurity.  The hopping matrix is given by $t_{ij} = 
t(|\bvec{R}_i - \bvec{R}_j|),$ where $t(r) = 2 (1 + r/a_B)e^{-r/a_B} \, {\rm Ry}$, \cite{Bhatt1}
and the Rydberg (Ry) is the binding energy of a hole.  We assume that the Mn spins are 
strongly localized and hence the exchange integral is given by $J_{ij} = J_0 \left|\phi_{ij}
\right|^2 = 
J_0 \, e^{-2|\bvec{R}_i - \bvec{R}_j|/a_B},$ which is proportional to the charge density at the 
$i^{th}$ Mn site of the carrier in the hydrogenic wavefunction around the $j^{th}$ Mn site.
The external magnetic field is given by $\bvec{H}$,
the Land\'{e} g-factors for the Mn spins and carriers are $g$ and $g^*$ respectively, 
and $\mu_B$ is the Bohr magneton.  Our simulations are for zero magnetic field.

We consider finite size cubic samples with $L$ cubic unit cells of the GaAs structure per side, 
with periodic boundary 
conditions, and $N_d$ Mn impurities.  Thus $x$ is given by $N_d/(4L^3)$, and is related to the 
concentration of Mn impurities, $n_{Mn}$  through $n_{Mn} = 4x/a^3$, where $a = 5.65$~\,\AA $\;$ 
is the GaAs lattice constant. 
The number of carriers is $N_h = p N_d$, with $p$ between
0.1 and 0.3, as indicated by experimental studies of samples grown by molecular beam epitaxy. 
\cite{Beschoten,Omiya} As mentioned previously, these low
values of $p$ are due to compensation processes, in which As antisites are believed to play an
important role. \cite{SHill}

The only difference between our model and that in Ref. \onlinecite{Mona1,Mona2}, 
is that instead of studying spin-$\frac{5}{2}$ Mn spins, we treat the Mn spins as 
classical Heisenberg vector spins.  This should be a reasonable approximation since $S = 5/2$ 
is a large spin and a Quantum Monte Carlo calculation would not gain much
due to the uncertainties in the materials parameters.  
[At the mean field level, this approximation
has the effect of lowering $T_c$ by a factor of $1 + \frac{1}{S}$ relative to the 
$T_c$ for quantum spins].  One quantity which differs significantly with classical spins is
the specific heat at low temperatures, \cite{Xin1,Xin2} which we have not studied. 

The values we use for numerical parameters are
$a_{B}$ = 7.8 \AA, 1 Ry = 112.4 meV, and $J_0$ = 15 meV, as discussed in
Ref. \onlinecite{Mona2}.  With these parameters we have an impurity band whose bottom lies around 200-300 meV
(2-3 Ryd) below the host conduction band and the Fermi energy varies from around 13 to 55 meV (depending on $x$ and $p$)
above the bottom of the impurity band
for the parameter range considered. These values are of the same order of
magnitude as the observed splitting and bandwidth of the impurity band in angle resolved photoemission
experiments. \cite{ARPES}

Our goal in this work is partly to 
understand the deviations from mean-field theory in the model Eq. (\ref{eq:Ham}), therefore
we also perform mean field calculations using Langevin functions to represent 
the polarization of classical spins (in the quantum case one uses Brillouin functions),
with which we compare our Monte Carlo results.  

\subsection{Method of Simulation}
Consider a system with classical spin degrees of freedom and fermionic 
degrees of freedom, such as described by the Hamiltonian (\ref{eq:Ham}).   
The assumption of classical spins means that one can parameterize the 
spin at each site by its $z$-component and azimuthal angle, i.e. 
$\bvec{S}_i = (S^z_i,\phi_i)$, and 

\begin{eqnarray}
S^x_i & = & S \sqrt{1 - \frac{{S^z_i}^2}{S^2}} \cos\phi_i, \\
S^y_i & = & S \sqrt{1 - \frac{{S^z_i}^2}{S^2}} \sin\phi_i.
\end{eqnarray}
For any given configuration of classical
spins, $\{\bvec{S}\} = \{S^z,\phi\}$, the Hamiltonian (\ref{eq:Ham}) can be 
diagonalized to give

\begin{equation}
{\mathcal H}(\{\bvec{S},a_n^\dagger,a_n\}) = \sum_n E_n(\{S^z,\phi\}) a^\dagger_n a_n,
\end{equation}
where the states $n$ are a diagonal basis
% for the Hamiltonian, 
and 

\begin{equation}
a^\dagger_{n} = \sum_{i\sigma} \psi_{n\sigma}(i) c^\dagger_{i\sigma}, \quad\quad a_{n} 
= \sum_{i\sigma} \psi^*_{n\sigma}(i) c_{i\sigma},
\end{equation}
are linear combinations of the $c^\dagger_{i\sigma}$ and $c_{i\sigma}$ operators. Here 
$\psi_{n\sigma}(i)$ is the eigenfunction for the $n^{\rm th}$ energy level at site $i$ for
spin $\mbox{\boldmath$\sigma$}$. Using the grand-canonical ensemble, the partition function is 

\begin{eqnarray}
\label{eq:GCPF}
Z & = & \left[\prod_{i=1}^{N_d} \int_{-1}^1 dS^z_i \int_0^{2\pi} d\phi_i \right] \sum_{n_1 = 0,1} \ldots
\sum_{n_{N_0}=0,1} \nonumber \\
& & \hspace*{1cm} \times 
 e^{-\beta\sum_{k=1}^{N_0} (E_n(\{S^z,\phi\}) - \mu)n_k},
\end{eqnarray}
where $n_k = 0,1$ is the occupation number of level $k$ (there are $N_0 = 2N_d$ levels), and
$\beta = 1/(k_B T)$ is the inverse temperature.  
Summing over fermion degrees of freedom leads to the result 

\begin{eqnarray}
Z & = & \left[ \prod_{i=1}^{N_d} \int_{-1}^1 dS^z_i \int_0^{2\pi} d\phi_i \right]
\prod_{n=1}^{N_0}
\left(1 + e^{-\beta(E_n(\{S^z,\phi\}) - \mu)}\right). \nonumber \\
\end{eqnarray}
This can be cast in a similar form to that used for a spin-only system 

\begin{eqnarray}
\label{eq:MetroF}
Z = \left[\prod_{i=1}^{N_d} \int_{-1}^1 dS^z_i \int_0^{2\pi} d\phi_i \right]\,
e^{-\beta {\mathcal F}_c(\{S^z,\phi\})},
\end{eqnarray}
where the carrier free energy ${\mathcal F}_c(\{S^z,\phi\})$ 
for a given configuration of classical spins is 

\begin{eqnarray}
\label{eq:Fcarr}
{\mathcal F}_c(\{S^z,\phi\}) = -\frac{1}{\beta} \sum_{n=1}^{N_0} \log\left(1 + 
e^{-\beta(E_n (\{S^z,\phi\}) - \mu)}\right).
\end{eqnarray}
The chemical potential $\mu$ is determined from the condition

\begin{eqnarray}
\left<N_c\right> & = & \frac{\partial}{\partial (\beta\mu)} \log Z , \nonumber \\
& = & \frac{1}{Z} \left[\prod_{i=1}^{N_d} \int_{-1}^1 dS^z_i \int_0^{2\pi} d\phi_i \right]
\sum_{n=1}^{N_0} f(E_n) e^{-\beta {\mathcal F}_c(\{S^z,\phi\})}, \nonumber \\
\end{eqnarray}
where 

\begin{equation}
f(E_n) = \frac{1}{e^{\beta(E_n -\mu)} + 1},
\end{equation}
is the Fermi distribution function and
$\left<N_c\right>$ is the expectation value for the number of carriers.

Equations (\ref{eq:MetroF}) and (\ref{eq:Fcarr}) imply that we can use Monte Carlo 
techniques to evaluate various thermodynamic 
quantities, except that we must use the carrier free energy rather than the 
internal energy in the Metropolis algorithm.  
An important point to note is that since the calculation is in the grand canonical
ensemble, both the temperature and chemical potential {\it must be kept fixed}
 during the simulation.
This differs from a recent Monte Carlo study\cite{Sakai} in which $N_c$ was held constant at each 
Monte Carlo step by varying $\mu$ rather than holding it constant, 
in our simulations we hold $\beta$ and $\mu$ fixed for each run and then 
we average over disorder using samples with equal $\left<N_c\right>$, since we wish to 
average samples which have the same $x$ and $p$.

\subsection{Magnetic and Thermodynamic Quantities}
\label{sec:MTQ}

Our Monte Carlo simulations allow us to compute various magnetic and thermodynamic quantities after equilibration.
We perform equilibrium thermal averages (indicated by $\left<\ldots\right>$) for each sample (realization
of disordered  Hamiltonian)  and then average 
over many different realizations of disorder. (The disorder average is indicated by an 
overbar $\overline{\cdots}$).  We collect data for both global (bulk) and local quantities.  We first consider the global
quantities we study. We compute the moments of the average magnetization per Mn 
spin, $M^q$, and the average magnetization per carrier, $m^q$, at each Monte Carlo step,
where the $q^{\rm th}$ moments are given by

\begin{eqnarray}
M^q & = & \left(\frac{\sqrt{|\sum_i \bvec{S}_i|^2}}{SN_d}\right)^q, \\
m^q & = & \left(\frac{1}{2}\frac{\sqrt{\left|\sum_{i,\alpha\beta} 
c^\dagger_{i\alpha}\mbox{\boldmath $\sigma$}_{\alpha\beta}
c_{i\beta}\right|^2}}{N_h}\right)^q .
\end{eqnarray}
In the above equation  $S = \frac{5}{2}$.
The average spin per Mn, $S_{Mn}$, and the average fermion spin, $s_c$ are thus

\begin{equation}
S_{Mn} = \overline{\left<M\right>}, \quad\quad s_c = -\overline{\left<m\right>} .
\end{equation}
{\em Note that with our definition, $S_{Mn}$ is normalized, i.e. for a fully polarized spin state $S_{Mn} = 1$}.
The negative sign for the fermion magnetization is 
due to the antiferromagnetic interactions between Mn spins and carriers which 
leads to oppositely oriented polarizations.
The Mn susceptibility is given by

\begin{equation}
\label{eq:mnchi}
\chi_{Mn} = \beta\overline{\left[\left<M^2
\right> - \left<M\right>^2\right]},
\end{equation} 
while the carrier susceptibility is
\begin{equation}
\label{eq:hchi}
\chi_h = \beta\overline{\left[\left<m^2
\right> - \left<m\right>^2\right]}.
\end{equation}
For each measured quantity, we calculate the statistical errors in the conventional manner 
from the variance.  One of the limitations of studying a magnetic model on small 
lattices, as we are forced to do here, is that it is not  easy to identify 
the position of a thermodynamic transition from considering
quantities such as the magnetization.  However, from  finite size
scaling theory, quantities that are dimensionless are particularly useful to identify
the transition temperature. One such quantity, known as the Binder cumulant,
is given by 
\begin{equation}
\label{eq:gl}
G(L,T) = \frac{1}{2}\left(5 - 3 \overline{\left(\frac{\left<M^4\right>}{
\left<M^2\right>^2}\right)}\right).
\end{equation}
$G(L,T)$, which measures the ratio of the fourth moment to the square 
of the second moment of the magnetization for a finite system, is defined such that in the paramagnetic phase
$G(L,T)$ decreases with $L$, and tends to zero as $L \rightarrow \infty$, while in the
ferromagnetic phase it increases with increasing size $L$ and tends to unity in the
thermodynamic limit. Near the transition temperature $T_c$, being dimensionless, it has the
finite size scaling form $G(L,T) = G[L^{{1 \over \nu}}(T-T_c)]$, \cite{Xin1,Xin2,BinderY} 
where $\nu$ is the exponent of the diverging spin-spin correlation length $\xi \sim (T - T_c)^{-\nu}$.
Consequently, at $T_c$, 
$G(L,T_c)$ is independent of $L$; $T_c$ can be identified by a simultaneous crossing of $G(L,T)$ vs. $T$ curves for
different $L$. Because $G$ is dimensionless, and depends only on the ratio $L/\xi$, rather than both $L$
and $\xi$,
this method is found to be more reliable in determining $T_c$ than analysis of the onset of 
magnetization, or peaks in the magnetic susceptibility in finite sized samples.

Local quantities we calculate are the 
local charge density  at each Mn site:

\begin{equation}
\label{eq:charge}
\rho_i = \sum_{j,\sigma} |\phi_{ij}|^2 \left<c^\dagger_{j\sigma}c_{j\sigma} \right>,
\end{equation}
and the local magnetization, which we define to be the average projection of the spin at
site $i$ on the 
total magnetization $\bvec{M}$

\begin{equation}
\label{eq:localm}
M^{\rm local}_i = \left<\bvec{S_i}\cdot\bvec{M}\right>.
\end{equation}
We are interested in the individual distributions ${\cal P}(\rho_i)$ and 
${\tilde{\cal P}}(M^{\rm local}_i)$ and the joint distribution 
${\cal P}_J(\rho_i,M^{\rm local}_i)$ to characterize the
local environment at different Mn sites.

\section{Perturbative Monte Carlo (PMC)}
\label{sec:PMC}
In principle, to perform Monte Carlo simulations on a model with fermion and classical degrees
of freedom, one needs to diagonalize the fermion part of the Hamiltonian after each ``spin
flip'' (more precisely a spin rotation), i.e. a new choice of the classical variable.
This leads to new eigenvalues which are used to compute the change in the carrier free energy
${\mathcal F}_c(\{S^z,\phi\})$.  
This is computationally time consuming, 
and hence one would prefer a quicker, approximate method which is still 
reasonably accurate.  One such approach is the 
hybrid Monte Carlo (HMC) algorithm used recently on the DMS problem, \cite{Schliemann} 
and also on the double exchange model.
\cite{HMC}  We have developed an algorithm that works in a similar spirit to HMC, that we 
describe below. 

We select a spin, at site $i$ and allow it to perform a small rotation:

\begin{eqnarray}
S^z_i & \to & S^z_i + \delta S^z_i, \\
\phi_i & \to & \phi_i + \delta \phi_i,
\end{eqnarray}
where $\delta S^z_i \in [-\lambda/2,\lambda/2]$, and 
$\delta\phi_i \in [-\lambda\pi,\lambda\pi]$,
are restricted to a small region on the surface of the unit sphere ($\lambda \ll 1$), in a way 
that leads to uniform sampling.  We use perturbation theory to compute the change in the carrier
eigenenergies due to the rotation of the spin at site $i$:

\begin{equation}
E_n(\{S^z,\phi\})  \to E_n(\{S^z,\phi\}) + \delta E_n
\end{equation}
where

\begin{equation}
\delta E_n = \frac{1}{2} \delta\bvec{S}_i \cdot \sum_j J_{ij}
\sum_{\alpha\beta} \psi^*_{n\alpha}(j) \mbox{\boldmath $\sigma$}_{\alpha\beta}
   \psi_{n\beta}(j).
\end{equation}
As a result, $\delta E_n \sim O(\lambda)$ and therefore perturbation theory is essentially 
exact as $\lambda \to 0$.

We now use Eq. (\ref{eq:Fcarr}) to compute the change in carrier free energy 
associated with the spin rotation, and the Metropolis criterion to decide whether to accept the
spin rotation.  We perform such an update for each spin in the system
using the perturbation scheme. After a complete sweep through the system, we 
compute the eigenenergies $E_n$ and eigenfunctions 
$\psi_n(i)$ corresponding to the new spin configuration
using exact diagonalization, and start a new Monte Carlo sweep.
We found that this approach was quicker than if we diagonalized after every spin flip, 
generally by a factor of 3-4.

\subsection{The chemical potential}
One important issue is the choice of  the chemical potential $\mu$ for the 
desired average number of charge carriers, $\left<N_c\right> = N_h$.
To determine the chemical potential we consider two replicas of the system, one starting from
a fully polarized (ferromagnetic) configuration, the other a purely random (paramagnetic) 
configuration.  After every few Monte Carlo steps (in practice 5 MC steps worked well), we use the condition 

\begin{equation}
N_h = \sum_{n} \frac{1}{e^{\beta(E_n - \mu)} + 1},
\end{equation}
to update the value of $\mu$ for each replica.  When the magnetization $M$ and chemical potential
of each replica agree to within 2 \%, we continue to calculate the chemical potential for each
replica after every 5 Monte Carlo steps, but their dynamics are determined by the mean chemical potential,
$\bar{\mu}$.
This average chemical potential is free to vary up until some equilibration time, and then the 
chemical potential is fixed as $\mu = \left<\bar{\mu}\right>$ where the average is over the time 
during the
equilibration period for which $\bar{\mu}$ is used to calculate the carrier free energy.  
The equilibration time used depended on the temperature, $x$ and $p$,
but was generally between 20000 and 40000 Monte Carlo steps. (The magnetization generally 
equilibrated within 2000 Monte Carlo steps, while the remainder of the equilibration was required to 
obtain an accurate value of the chemical potential).  The fixed chemical potential is used for
the remainder of the run, during which data is collected.  We typically use 20000 to 40000
Monte Carlo steps to collect data.  
This procedure was found to obtain a chemical potential that yielded values of $\left<N_c\right>$
for the sample that were generally within 2\% of $N_h$
(not surprisingly it was found to be more effective in larger samples where the 
relative sizes of the fluctuations are correspondingly smaller).

In several other studies of models with fermions coupled to spins 
\cite{Tc,Schliemann,HMC} the electron occupation numbers
have been taken to be those corresponding to $T=0$ (i.e. the Fermi distribution is 
replaced by a Heaviside distribution).
In our simulations, we allow the filling to change as a function of temperature (i.e. we 
do not assume degenerate electrons).  We do, however 
truncate the number of states we include -- their number varies as a function of temperature, 
such that states of high energy which have $10^{-5}$ or lower 
filling are discarded (the cutoff is sufficiently small that the results are not sensitive
to its value).  

\subsection{Testing the PMC algorithm}
\begin{figure}[htb]
\centerline{\psfig{file=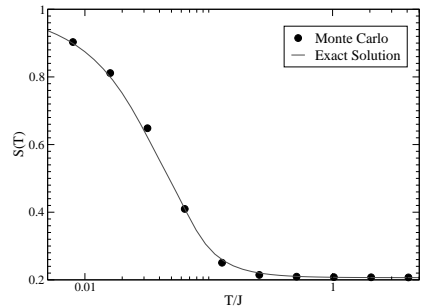,height=4.5cm,angle=270}}
\caption{Comparison of exact result (line) and Monte Carlo (points) for the magnetization of the toy model, 
for $t=J$, $N = 20$ and $N_h = 3$. 40000 Monte Carlo steps were used to generate the MC results.}
\label{mag1}
\end{figure}

We tested our perturbative Monte Carlo algorithm on a simple model of fermions coupled to classical spins 
which admits an exact solution.
The Hamiltonian, shown below in Eq.~(\ref{eq:Mona}), corresponds to fermions hopping
along a 1-dimensional chain with $N$ sites and periodic boundary conditions.  At every 
lattice site $i$, there is a classical spin $\bvec{S}_i$, and the fermions are equally strongly
coupled to all spins on the chain. 
The model Hamiltonian is thus

\begin{eqnarray}
\label{eq:Mona}
{\mathcal H} & = & t\sum_{\left<ij\right>,\sigma} c^\dagger_{i\sigma} c_{j\sigma} + \frac{J}{N}
\sum_{i}\bvec{S}_i \cdot \sum_{j,\alpha\beta}c^\dagger_{j\alpha} 
\frac{1}{2}\mbox{\boldmath${\sigma}$}_{\alpha\beta} c_{j\beta},
\end{eqnarray}
where energy is measured in units of $t=1$ and only nearest neighbour hopping 
is allowed.  Note that the exchange  scales as $J/N$ to ensure an extensive energy.
The eigenvalues for this model can be calculated exactly as discussed in Appendix~\ref{app:toy}.
Exact
results for any number of classical spins, $N$, and $N_h$ fermions can be calculated, 
particularly 
for the magnetization per spin, $S$, the Binder cumulant $G(L)$ and the fermion
magnetization $s_c$. A comparison of the exact results for $S(T)$ and those 
calculated using PMC are shown in Fig.~\ref{mag1} for $t=J$, $N=20$, $N_h = 3$ and using 40000
Monte Carlo steps.

\begin{figure}[htb]
\centerline{\psfig{file=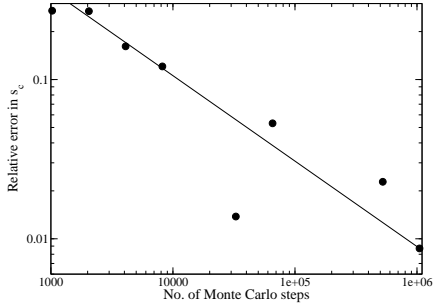,height=4.5cm,angle=270}} \vspace{0.2cm}
\caption{Convergence of the fermion magnetization for the toy model with the same parameters
as in Fig. \ref{mag1}, at termperature $T/J = 0.0064$.  
The straight line has a $1/\sqrt{t}$ dependence, where $t$ is the number of Monte Carlo steps.}
\label{sqrtt}
\end{figure}

In Fig. \ref{sqrtt} we plot the relative error in the fermion magnetization against the number 
of Monte Carlo steps $(t)$ for the same parameters used for Fig.~\ref{mag1} at $T/J = 0.0064$ 
where the magnetization
is about $\approx 0.5$. At long times the convergence is of the form $t^{-\frac{1}{2}}$,
where $t$ is the number of Monte Carlo steps, as expected for Monte Carlo simulations.
We have also made a comparison between the PMC method and diagonalization after every spin rotation 
for the model of DMS that we are really interested in [Eq. (\ref{eq:Ham})], 
which gave agreement to within the 2 \% statistical error bars. 
There is also some contribution to the systematic error 
from our approach, but we choose $\lambda$ to be small (generally $\lambda \leq 0.03$), 
so that this error has little effect on our results.

\section{Results: Global Quantities}
\label{sec:results}
Four different values of $x$ and $p$ were chosen for the Monte Carlo study,
representative of the concentrations and compensation seen in the experimental materials.
  These were:
$x = 0.01$, $p = 0.1$; $x =0.01$, $p = 0.3$; $x = 0.03$, $p = 0.1$ and $x =0.03$, $p = 0.3$. 
Note that using Mott's criterion as in Ref. \onlinecite{Mona2}, the $x=0.01$, $p = 0.1$ 
sample is 
in the insulating phase, whilst the remainder of the cases considered are somewhere on the 
metallic side of the MIT. In the case of the most metallic sample $x=0.03$, $p=0.3$ it may be 
inappropriate to neglect the 
host band states for a quantitative description of the magnetic properties, 
but we have included this also within the context of an impurity 
band model for comparison with the other cases. 

\begin{table}[htb]
\begin{tabular}{c|c|ccc|c}
$x$ & $p$ & $L$ & $N_d$ & $N_h$ & Case \\ \hline
0.01 & 0.1 & 11 & 53 & 5 & A1\\
 & & 12 & 69 & 7 & A2\\
 & & 14 & 110 & 11 & A3\\
0.01 & 0.3 & 11 & 53 & 16 & B1 \\
 & & 12 & 69 & 21 & B2\\
0.03 & 0.1 & 7 & 41 & 4 & C1\\
 &  & 8 & 61 & 6 & C2 \\
0.03 & 0.3 & 7 & 41 & 12 & D1 \\
 & & 8 & 61 & 18 & D2
\end{tabular}
\caption{Sample sizes considered for different values of $x$ and $p$.}
\label{tab:one}
\end{table} 

We carried out simulations on lattices of linear size between $L = 7$ and $L = 14$, which 
contained between $40$ and $110$ Mn spins and between $4$ and $21$ carriers.
The sizes that were considered and their labels are tabulated in Table \ref{tab:one}.
We averaged up to 700 samples per data 
point, depending on the size and temperature. Typically at least 30-40 samples were averaged 
for each data point.  In this section we present our results for global quantities such as 
the magnetization, the Binder 
cumulant (to determine $T_c$) and the magnetic susceptibility. We also present our results 
comparing ordered and disordered samples. We focus on local 
quantities in the next section. 

\subsection{Magnetization}

\begin{figure}[htb]
\centerline{\psfig{file=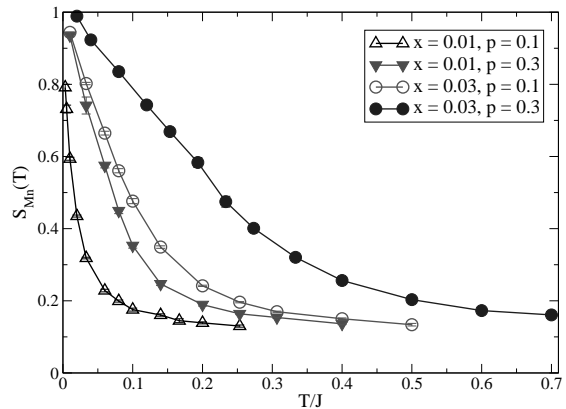,height=6cm,angle=270}}
\caption{Mn magnetization as a function of temperature (normalized to saturation) for case A2 ($x=0.01$, $p=0.1$),
case B2 ($x=0.01$, $p=0.3$), case C2 ($x=0.03$, $p=0.1$), 
 and case D2 ($x=0.03$, $p=0.3$).}
\label{Mnmag}
\end{figure}

In Fig.~\ref{Mnmag} we plot the average magnetization per Mn spin, $S_{\rm Mn}(T)$, as a function 
of temperature for all four combinations of $x$ and $p$ using samples containing between 
60 and 70 Mn spins in each case (thus finite size effects are similar in all four curves).
 Clearly, the critical temperature, $T_c$ increases
with increasing $x$ and $p$. The magnetization curves for cases B2 and C2, which have the same hole concentration $px$ 
have similar numerical values, although the curve for $x=0.01$,
$p=0.3$ appears to have a lower $T_c$.  The most important feature is the unusual shape of the 
magnetization curves. The magnetization decreases rapidly from full polarization at $T=0$, leading to 
linear or concave upward shapes, similar to those found in the mean field 
approximation \cite{Mona1,Mona2} and for insulating II-VI DMS, \cite{Xin1,Xin2} and very 
unlike the strongly convex upward magnetization curves seen in conventional ferromagnets such as iron. 
\cite{Review}

\begin{figure}[htb]
\centerline{\psfig{file=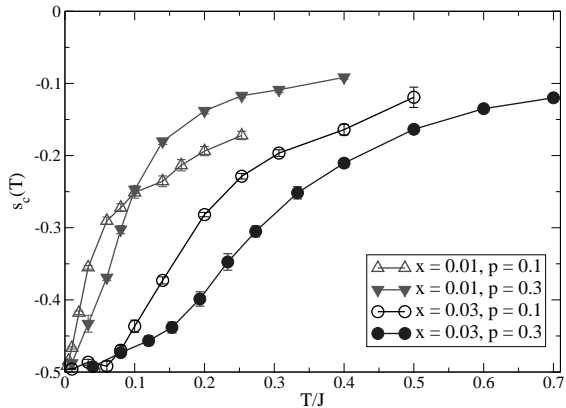,height=6cm,angle=270}}
\caption{Carrier magnetization as a function of temperature for case A2 ($x=0.01$, $p=0.1$),
case B2 ($x=0.01$, $p=0.3$), case C2 ($x=0.03$, $p=0.1$), 
 and case D2 ($x=0.03$, $p=0.3$).}
\label{hmag}
\end{figure}

The carrier magnetizations, $s_c (T)$ for the same sizes shown in  Fig.~\ref{Mnmag} are
shown in Fig.~\ref{hmag}.  The curves mirror the feature of the Mn magnetizations that increasing
$x$ and $p$ lead to greater polarization at higher temperatures -- there is also a clearer 
distinction between curves for B2 and C2, where $x = 0.01$, $p =0.3$ polarizes at noticeably lower 
temperatures than $x = 0.03$, $p = 0.1$.  One major 
difference from the results of the mean field study is that the curves for the carrier 
magnetizations appear to be much more like the Mn magnetization curves, 
whereas in the mean field study \cite{Mona1,Mona2} the carrier curves remained almost fully 
polarized until $T$ was close to $T_c$.  This feature is in closer agreement with experiments than the
mean field results. \cite{Beschoten}

\subsection{Critical Temperature}
\begin{figure}[htb]
\centerline{\psfig{file=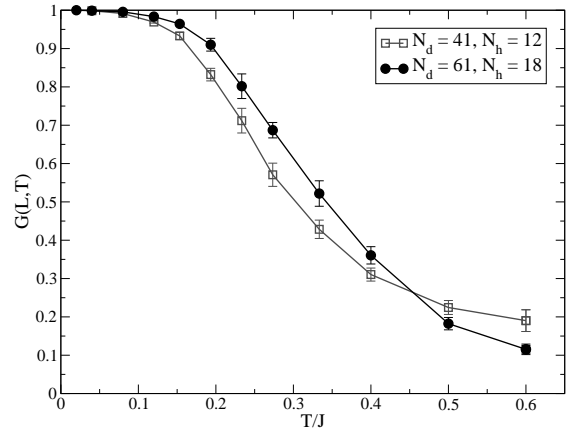,height=6cm,angle=270}}
\caption{Binder Cumulant as a function of temperature for $x = 0.03, p=0.3$.  The data shown is 
for cases D1 ($N_d = 41$, $N_h = 12$) and D2 ($N_d = 61$, $N_h = 18$).}
\label{gl33}
\end{figure}

We use the Binder cumulant curves, $G(L,T)$ (see Eq.~(\ref{eq:gl})) to identify 
the critical temperature $T_c$ for each carrier and Mn concentration.  As described
in Sec. \ref{sec:MTQ}, $T_c$ is indicated by the simultaneous intersection of
$G(L,T)$ curves for sufficiently large sizes, $L$.  Fig. \ref{gl33} shows $G(L,T)$ 
curve for $x = 0.03$ and $p = 0.3$ for both sample sizes (D1 and D2 of Table \ref{tab:one})
over a wide temperature range ($T/J$ = 0.02 to 0.6).  As expected $G(L,T)$ 
decreases with increasing size $L$ at high $T$ whereas the reverse is true at 
low $T$.  At the transition temperature $T_c$, $G(L,T)$ is expected to be 
independent of $L$, which is indicated by the crossing point of the two curves, implying
$T_c/J = 0.45$.  The solid curves are spline fits to the data appropriately weighted by 
the error bars.  

\begin{figure}[htb]
\centerline{\psfig{file=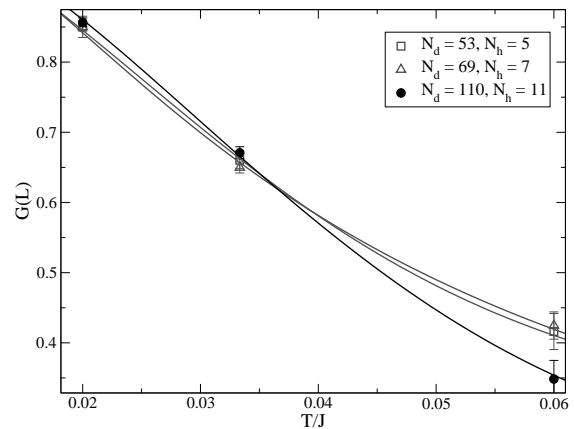,height=6cm,angle=270}}
\caption{Binder Cumulant as a function of temperature for $x = 0.01, p=0.1$.  The data shown is 
for cases A1 ($N_d = 53$, $N_h = 5$), A2 ($N_d = 69$, $N_h = 7$) and A3 ($N_d = 110$, $N_h = 14$).  
The curves for each size are obtained from a non-linear curve fit using the entire temperature range.}
\label{gl11}
\end{figure}

\begin{figure}[htb]
\centerline{\psfig{file=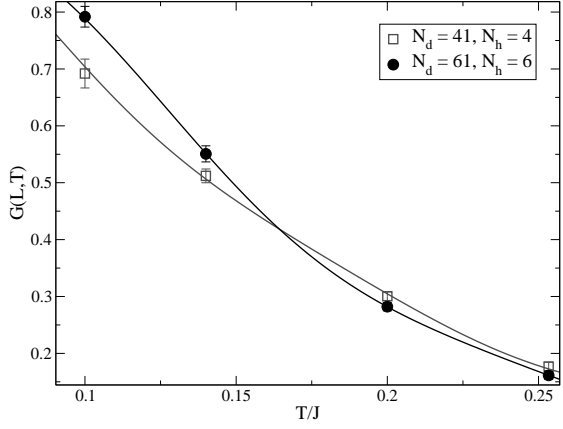,height=6cm,angle=270}}
\caption{Binder Cumulant as a function of temperature for $x = 0.03, p=0.1$.  The data shown is 
for C1 ($N_d = 41$, $N_h = 4$) and C2 ($N_d = 61$, $N_h = 6$). The curves for 
each size are obtained from a non-linear curve fit using the entire temperature range. Up to 
700 samples were averaged for the data shown.}
\label{gl31}
\end{figure}

\begin{figure}[htb]
\centerline{\psfig{file=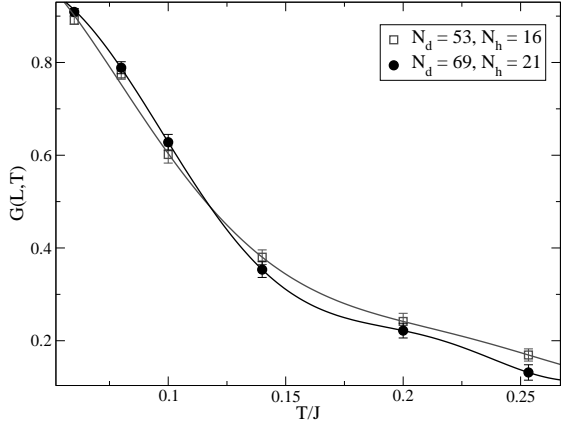,height=6cm,angle=270}}
\caption{Binder Cumulant as a function of temperature for $x = 0.01, p=0.3$.  The data shown is 
for $N_d = 53$, $N_h = 16$ and $N_d = 69$, $N_h = 21$. The curves for 
each size are obtained from a non-linear curve fit using the entire temperature range.}
\label{gl13}
\end{figure}

Figs. \ref{gl11} - \ref{gl13} show the $G(L,T)$ data for the other three 
concentrations studied, each in the vicinity of the crossing point.  Several hundred 
samples were generally averaged for each data point as indicated in the figure captions.  {\it The 
solid curves are spline fits to the data over the entire temperature range studied (typically 
$T/T_c$ = 0 to 3).}  For the one case where we studied more than two system sizes ($x = 0.01$, 
$p = 0.1$), which we exhibit in Fig. \ref{gl11}, all curves are consistent with a single
intersection point, although a small size dependence cannot be ruled out. \cite{footnote}
Such a dependence would represent corerections to finite size scaling, arising from the 
relatively small sizes of the samples studied.  (The effective linear size of the spin 
system is $N_d^\frac{1}{3}$, which varies from 3.4 to 5). The small sizes, necessitated 
by the need to repeatedly diagonalize the fermion Hamiltonian, also limited the dynamic
range available in our study.  Because of this, the curves for the sizes studied do not 
splay out very dramatically around $T_c$.  This, in turn, necessitates calculation of 
$G(L,T)$ to high precision, which requires long runs and averaging over many samples.
Despite these drawbacks, we find $G(L,T)$ to be a more reliable estimator of $T_c$ than
e.g. peaks in $\chi(T)$ for the largest size studied.  Based on Figures \ref{gl33} to 
\ref{gl13}, we estimate $T_c$ as shown in Table \ref{tab:tc}.

In Table \ref{tab:tc} we compare the Curie temperature $T_c$ as determined in the Monte Carlo
simulations here with the results of the mean-field approximation using Langevin 
functions for classical vector spins. 

\begin{table}[htb]
\begin{tabular}{c|c|ccc|c}
$x$ & $p$ & $T_c^{MC}/J$ & $T_c^{MC}/T_c^{MF}$  \\ \hline
0.01 & 0.1 & 0.037 $\pm$ 0.004 & 0.14 \\
0.01 & 0.3 & 0.12 $\pm$ 0.04 & 0.48 \\
0.03 & 0.1 & 0.16 $\pm$ 0.02 & 0.37 \\
0.03 & 0.3 & 0.45 $\pm$ 0.03 & 0.82 \\
\end{tabular}
\caption{Comparison of $T_c$ determined using Monte Carlo $(T_c^{MC})$ and mean field $(T_c^{MF})$.}
\label{tab:tc}
\end{table} 
As expected, the mean-field approximation overestimates $T_c$. 
While the reduction due to fluctuations is only about $20\%$ for the largest $x$ and $p$ studied, it can be
much more significant (a factor of 5 or more) at densities at and below the metal insulator transition 
density. (We mention in passing  that results in the mean-field approximation for quantum spins using Brillouin
functions are higher than for classical spins by $60-80\%$). 

Since there are many models for DMS which are
solved at the mean field level, having an understanding of the behaviour of these models 
when fluctuations are considered is very important if one wants to make quantitative 
fits to data.  For a  model where Mn ions interact with  carriers in an unperturbed host band, 
a similar reduction 
in $T_c$ was found in Monte Carlo simulations. \cite{Schliemann}  One important difference 
between the results here and those found in Ref. \onlinecite{Schliemann} however, is that here
mean field theory becomes more accurate with increasing carrier concentration at a given $x$,
whereas the opposite appears to be the case in that model.

The Monte Carlo simulations show clearly a strong decrease of $T_c$ at low carrier density 
$(px)$, in agreement with experiment, and rectify the unphysically large $T_c$ obtained in the
position dependent mean-field treatment in this limit. From Table \ref{tab:tc} it appears clear 
that the main dependence of $T_c$ comes from the carrier density $(px)$, with a relatively weaker 
dependence on $(x/p)$.

\subsection{Magnetic Susceptibility}
\begin{figure}[htb]
\centerline{\psfig{file=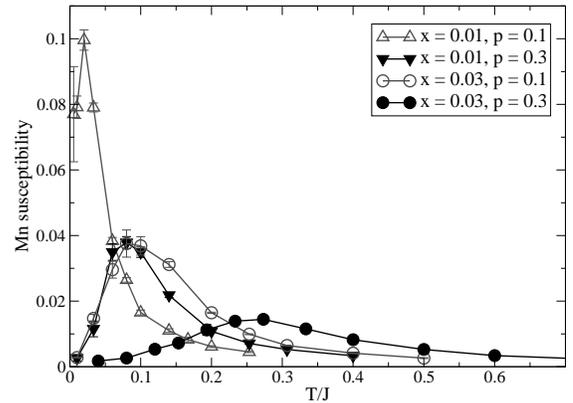,height=6cm,angle=270}}
\caption{Mn susceptibility per Mn spin as a function of temperature for 
for case A2 ($x=0.01$, $p=0.1$),
case B2 ($x=0.01$, $p=0.3$), case C2 ($x=0.03$, $p=0.1$), 
 and case D2 ($x=0.03$, $p=0.3$).}
\label{mnchi}
\end{figure}

\begin{figure}[htb]
\centerline{\psfig{file=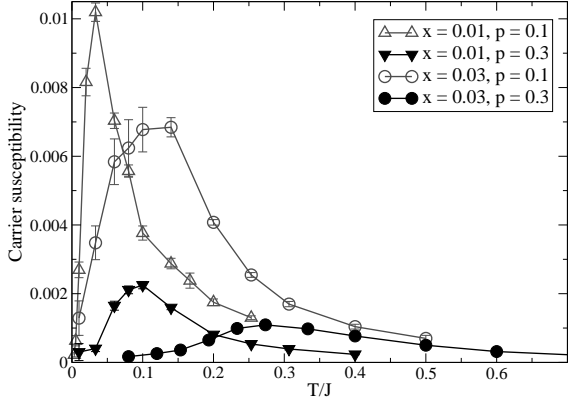,height=6cm,angle=270}}
\caption{Carrier susceptibility per carrier as a function of temperature for 
for case A2 ($x=0.01$, $p=0.1$),
case B2 ($x=0.01$, $p=0.3$), case C2 ($x=0.03$, $p=0.1$), 
 and case D2 ($x=0.03$, $p=0.3$).}
\label{hchi}
\end{figure}

The Mn and carrier susceptibilities [Eqs. (\ref{eq:mnchi}) and (\ref{eq:hchi})] are shown in Fig.~\ref{mnchi} and Fig.~\ref{hchi}
respectively, as a function of temperature, for the same sizes and dopings that were used for Mn and carrier magnetizations in 
Fig.~\ref{Mnmag} and Fig.~\ref{hmag}.  

At high temperatures, well above $T_c$, one expects 
\begin{equation}
\label{neweq}
\frac{1}{\chi} \simeq \frac{3k_B T}{n(g\mu_B)^2 S^2} \left( 1 - \frac{\theta}{T}\right),
\end{equation}
for classical spins (the factor of $S^2$ is replaced by $S(S+1)$ for quantum spins).  We have 
investigated the behaviour of the inverse susceptibility in the temperature range $T = 2\, T_c$ 
to $6\, T_c$, and found that the effective spin $S_{\rm eff} \simeq 1.6 - 1.7 \, S$, which 
suggests spin clusters (polarons) exist in the system well  above $T_c$.
Fitting data close to $T_c$ ($T_c$ to 2 $T_c$) Eq. (\ref{neweq}) gives $S_{\rm eff} \sim 2 S$, 
implying a significant
enhancement of the Curie constant over the high $T$ value. An even larger enhancement
has been seen in (Ga,Mn)N.\cite{private}

\subsection{Effects of Disorder}
One of the major results of the mean field study was that samples with maximal disorder in the
position of Mn spins were found to have a higher $T_c$ than those with less disorder. 
\cite{Mona1} This was
explained as being due to carriers being able to lower their total energy more in regions with 
higher Mn density, by polarizing Mn spins and then hopping between these sites.  However, it is
expected that the mean field factorization, which assumes that the carrier spin is either 
directed parallel or antiparallel to the overall magnetization, tends to align ``islands'' that
might not otherwise be aligned until lower temperatures.  This would be more likely to lead to 
a larger decrease in $T_c$ for disordered samples when thermal fluctuations are considered, 
since global magnetization around $T_c$ is more likely to be destroyed than in ordered samples.

We have tested whether the finding of disorder enhancing $T_c$ is robust, by investigating two
different cases.  The first is for $x = 0.01$, $p=0.1$, where we compared the weakly disordered
case and fully disordered case, and the second is for $x = 0.03$, $p=0.3$, where we compared the
fully ordered case and the fully disordered case.  

\begin{figure}[htb]
\centerline{\psfig{file=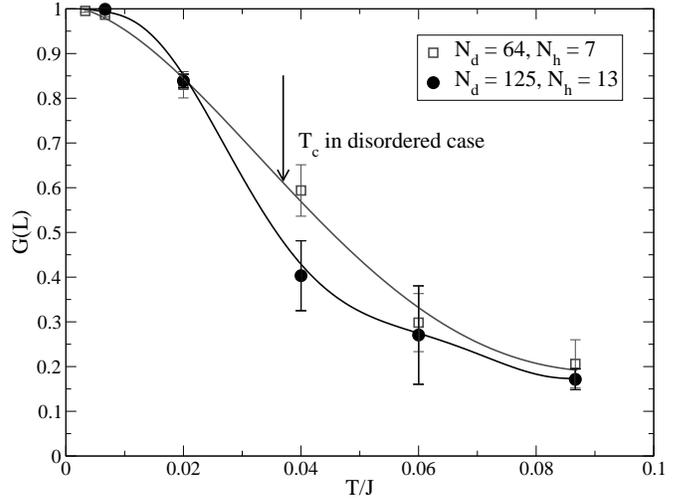,height=7cm,angle=270}}
\caption{Binder cumulant for $x \simeq 0.01$, $p \simeq 0.1$ for the mildly disordered case, 
shown for $N_d = 64$, $N_h = 7$ and $N_d = 125$, $N_h = 13$.}
\label{order11}
\end{figure}
In both cases we observed that the more ordered 
sample had a lower $T_c$.  However, the enhancement is much smaller than obtained within the mean field
treatment, as the arguments given above would suggest.

For $x = 0.01$, $p =0.1$, we determined the value of $T_c/J \simeq 0.037$ for the fully 
disordered case.  We define mild disorder to correspond to the situation where Mn spins are 
chosen to be displaced from a fully ordered Mn lattice to one of the 12 nearest neighbour sites 
on the fcc sublattice.  We considered $N_d = 64$, $N_h = 7$ and $N_d = 125$, $N_h = 13$, and 
then used the Binder cumulant to estimate $T_c$.  

\begin{figure}[htb]
\centerline{\psfig{file=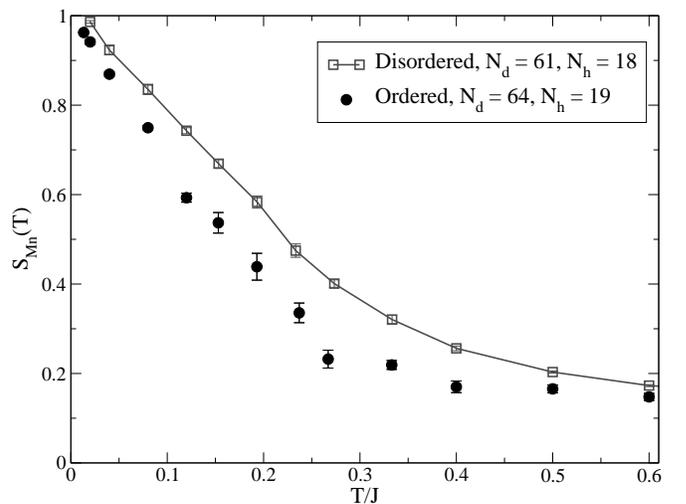,height=7cm,angle=270}}
\caption{Magnetization for $x \simeq 0.03$, $p \simeq 0.3$ for the completely disordered case, 
shown for D2 ($N_d = 61$, $N_h = 18$) and the fully ordered case $N_d = 64$, $N_h = 19$.}
\label{order33}
\end{figure}
Note that the number of Mn spins in our
simulations are 
slightly different from those used for other cases, since an ordered lattice of Mn needs to be 
commensurate with the underlying lattice. 
Whilst we do not have a particularly accurate
determination of $T_c$, it is clear in Fig.~\ref{order11} that the $G(L)$ curves cross at a 
temperature well below the $T_c$ of the disordered case ($T_c/J = 0.037$).

In Fig.~\ref{order33} we compare the magnetization calculated for $x = 0.03$, $p = 0.3$ with the
completely random case D2 and the completely ordered case with $N_d = 64$,
$N_h = 19$.  The $x$ and $p$ value were chosen since they give the largest $T_c$ and it was hoped that the 
effects would be more noticeable.  It is clear that even when thermal fluctuations are included,
the more ordered sample has a lower value of $T_c$.  (The magnetization at high $T$ is a finite size 
effect).

\section{Results: Local quantities}
\label{sec:five}

We now discuss our results for various local quantities such as the local magnetization and 
local charge densities for various Mn and carrier concentrations.

\subsection{Charge densities}
\begin{figure}[htb]
\centerline{\psfig{file=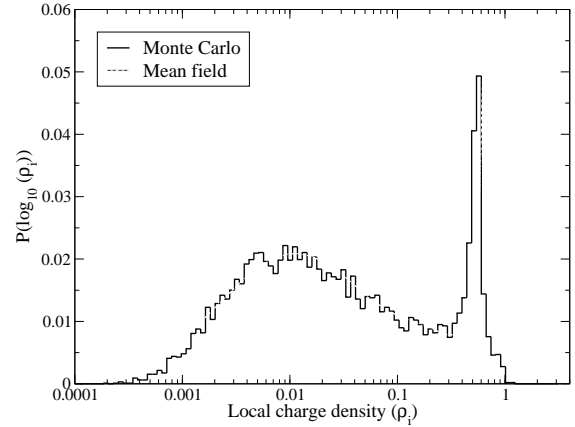,height=6cm,angle=270}}
\caption{Histogram of local charge densities calculated using Monte Carlo for 
$x = 0.01$, $p = 0.1$, $N_d = 110$, $N_h = 11$ (solid line), and the corresponding 
distribution calculated using 
mean field theory and Brillouin functions (Ref. 8).
The temperature is $T/J = 0.01 $ ($T_c/J \simeq 0.037$).}
\label{phist11}
\end{figure}

In Fig. \ref{phist11} we show a histogram of the local carrier charge densities $\rho_i$
[see Eq. (\ref{eq:charge})] at all Mn sites $i$, for $x = 0.01$, $p = 0.1$ (case A3) 
and averaged over 89 samples at the temperature $T/J = 0.01$, compared with the corresponding
distribution for $x = 0.0093$
and $p = 0.1$ obtained using the mean field approximation with Brillouin functions
close to $T = 0$. \cite{Mona2} Clearly the two distributions agree very well overall.
 The two peaks of the histogram suggest two populations of Mn sites.  
The peak at $\rho_i \simeq 0.7$ is from sites which have a high probability of trapping 
a carrier,
whilst the broader peak at much lower values of $\rho_i$ is due to sites which have very little
probability of having a carrier on them.  For these sites,  $\rho_i$
comes mainly through tailing from nearby sites that have a high charge carrier density 
[see Eq. (\ref{eq:charge})].  This shows that the charge carriers reside primarily on sites where 
there is a higher than 
average local Mn concentration, due to the strong interactions with these Mn spins.  
Because of the inhomogeneous charge distribution, spins of Mn atoms on sites devoid of charge carriers have very small 
antiferromagnetic couplings to the charge carriers, and therefore remain essentially
free down to low temperatures.  This explains the unusual  shape of 
the magnetization curves.

If we compare the histograms for the charge densities  with the data
obtained in the mean field model, we notice two important features.  Firstly, the histogram is
much less temperature dependent for $T \le T_c$.  This is probably because we are generally looking 
at lower temperature
scales than in the mean field case.  Secondly, these distributions have large widths typical of highly
disordered systems.  This large width was first found in the mean-field case, and
is the motivation for a simplified phenomenological model for DMS, based on a two-component
picture. \cite{MMR}   In this simplified model, one can divide the Mn spins into strongly and
weakly interacting components depending on the temperature.  This simplified model has been shown 
to be adequate to reproduce the results of the full distribution at the mean field level, and can also
explain experimental results on a qualitative basis.\cite{MMR}

\begin{figure}[htb]
\centerline{\psfig{file=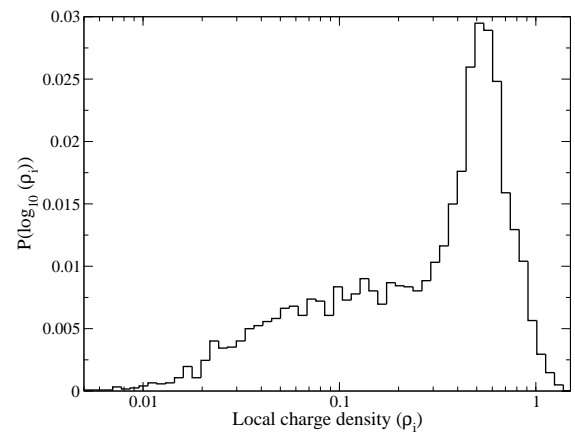,height=6cm,angle=270}}
\caption{Histogram of local charge densities for $x = 0.01$, $p = 0.3$, $N_d = 69$, $N_h = 21$. 
The temperature is $T/J = 0.1$ ($T_c/J \sim 0.12 $).}
\label{phist13}
\end{figure}

In Fig.~\ref{phist13} we show a histogram of the total charge densities $\rho_i$
 at all Mn sites $i$, for $x = 0.01$, $p = 0.3$, $N_d = 69$, $N_h = 21$
and averaged over 41 samples at temperature $T/J = 0.1$.  The most noticeable change  
as the temperature is varied is that the height of the peak at large $\rho_i$ 
decreases, corresponding to carriers becoming less localized at higher temperatures.

\begin{figure}[htb]
\centerline{\psfig{file=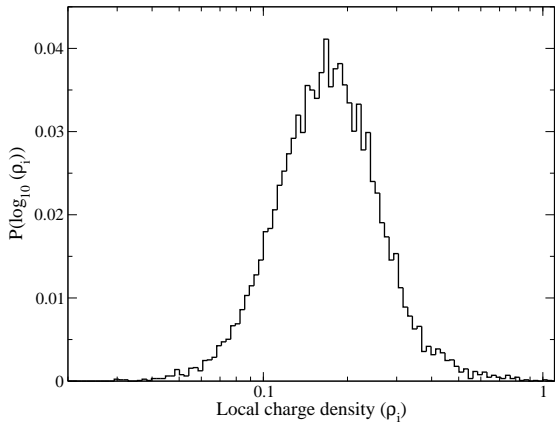,height=6cm,angle=270}}
\caption{Histogram of local charge densities for $x = 0.03$, $p = 0.1$, $N_d = 41$, $N_h = 4$. 
The temperature is $T/J = 0.14$ 
($T_c/J \simeq 0.16$).}
\label{phist31}
\end{figure}

\begin{figure}[htb]
\centerline{\psfig{file=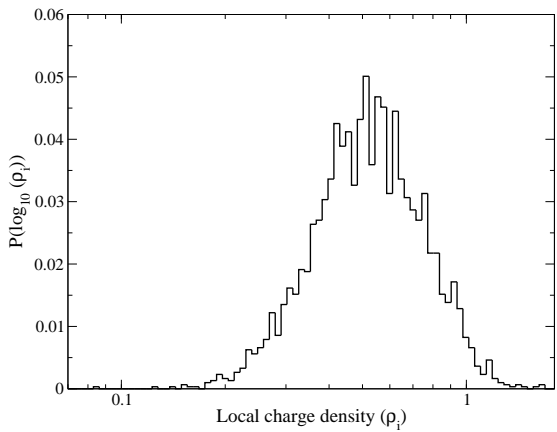,height=6cm,angle=270}}
\caption{Histogram of local charge densities for $x = 0.03$, $p = 0.3$, $N_d = 41$, $N_h = 12$. 
The temperature is $T/J = 0.4$  ($T_c/J \simeq 0.45$).}
\label{phist33}
\end{figure}

In Fig.~\ref{phist31} and \ref{phist33} we show the histogram of the total electron charge 
densities $\rho_i$
at each Mn site $i$, for $x = 0.03$, $p = 0.1$, $N_d = 41$, $N_h = 4$ at a temperature 
$T/J = 0.14$, averaged over 315 samples, and for $x=0.03$, $p=0.3$, $N_d=41$, $N_h = 12$  at 
a temperature $T/J  = 0.4$, averaged over 76 samples.  (In both cases the temperatures 
are around 90\% of $T_c$).   Unlike the case of lower $x$, these distributions have only one peak, 
however similarly to the lower $x$ case, the distribution is virtually independent of $T$ below $T_c$.

Besides the change in the shape of the distribution, the width of the distribution is also significantly
smaller at the higher hole concentration.  This is consistent with the observation that the eigenstates
at the Fermi energy are found to be delocalized at $x = 0.03$, whilst they appear to be localized at 
$x =0.01$. \cite{Mona1}

\subsection{Local magnetizations}
\begin{figure}[htb]
\centerline{\psfig{file=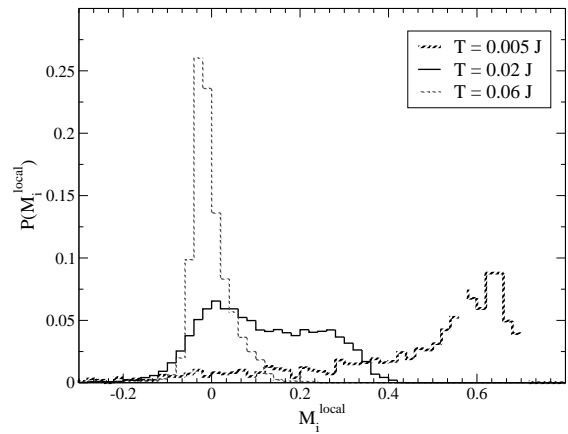,height=6cm,angle=270}} \vspace{0.2cm}
\caption{Histogram of local magnetizations for $x = 0.01$, $p=0.1$, $N_d = 110$, $N_h = 11$ at 
temperatures $T/J = 0.0053$, 0.02, and 0.06 ($T_c/J \simeq 0.04$).}
\label{mhist11}
\end{figure}

We now consider the local magnetizations.
In Fig.~\ref{mhist11} we show a histogram of the local magnetizations $M^{\rm local}_i$
[see Eq.~(\ref{eq:localm})] at all Mn sites $i$, for $x = 0.01$, $p = 0.1$, 
$N_d = 110$, $N_h = 11$  at three temperatures, one well below $T_c$, one around $T_c/2$ and one 
around $3T_c/2$. The number of samples to generate the histogram was 10 at $T/J = 0.0053$, 188 at 
$T/J = 0.02$ and 58 at $T/J = 0.06$. At the lowest temperature, there is a peak at
around $M_i^{\rm local} = 0.7$, and then a very broad tail that stretches to local magnetizations 
that are antiparallel to the overall magnetization.  For intermediate 
temperatures ($T = 0.02 \, J \simeq T_c/2$), 
there is evidence of two populations 
of Mn spins illustrated by the two peaks in the histogram, whilst for high temperatures (above $T_c$),
there is a peak centered very close to zero local magnetization with some weight for small 
local magnetizations.  
This is consistent with the two component picture described in the section on the local charge
distribution -- at temperatures well below $T_c$ most of the Mn spins are strongly coupled,  
leading to a peak and then a tail of more weakly coupled spins.  At intermediate temperatures,
there are two comparable populations of weakly and strongly coupled Mn spins, whilst at high 
temperatures, the local magnetizations are all small and there is no long-range magnetic order.

\begin{figure}[htb]
\centerline{\psfig{file=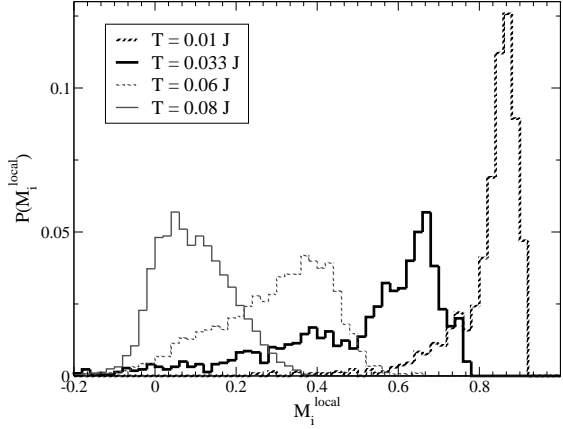,height=6cm,angle=270}}
\caption{Histogram of local magnetizations for $x = 0.01$, $p=0.3$, $N_d = 69$, $N_h = 21$ at 
temperatures $T/J = 0.01$, 0.033, 0.06, and 0.08
($T_c/J \simeq 0.12$).}
\label{mhist13}
\end{figure}

Fig. \ref{mhist13} shows a similar histogram of the local magnetizations $M^{\rm local}_i$,
for $x = 0.01$, $p = 0.3$, $N_d = 69$, $N_h = 21$ at temperatures $T/J = 0.01$, 
0.033, 0.06, and 0.08 averaged over 12, 20, 70 and 106
samples respectively.  The feature of two populations of Mn spins
 also appears to be present in this case, for $T/J$ between $0.03$ and $ 0.06$  ($T_c/J = 0.12$).

\begin{figure}[htb]
\centerline{\psfig{file=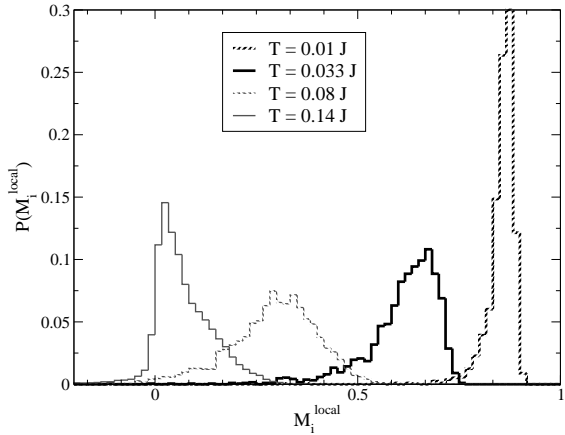,height=6cm,angle=270}}
\caption{Histogram of local magnetizations for $x = 0.03$, $p=0.1$, $N_d = 41$, $N_h = 4$ at 
temperatures $T/J = 0.01$, 0.033, 0.08, and 0.14 
($T_c/J \simeq 0.18$).}
\label{mhist31}
\end{figure}

\begin{figure}[htb]
\centerline{\psfig{file=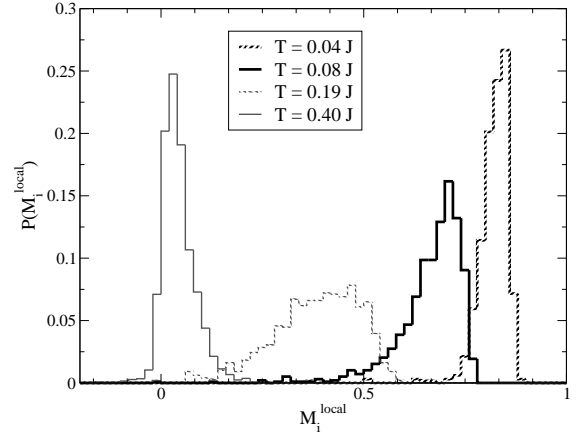,height=6cm,angle=270}}
\caption{Histogram of local magnetizations for $x = 0.03$, $p=0.3$, $N_d = 41$, $N_h = 12$ at 
temperatures $T/J = 0.04$, 0.08, 0.19, and 0.40  
($T_c/J \simeq 0.45$).}
\label{mhist33}
\end{figure}

Fig. \ref{mhist31} and \ref{mhist33} show the temperature evolution of the corresponding histograms
for $x = 0.03$, $p = 0.1$, ($N_d = 41$, $N_h = 4$) and $x = 0.03$, $p = 0.3$ ($N_d = 41$, $N_h = 12$) 
for temperatures ranging from well below to just below $T_c$.  For the former, with $T_c/J = 0.18$ 
the data are for $T/J = 0.01$, 0.033, 0.08, and 0.14, averaged over 49, 48, 48 and 334 samples
respectively, whilst for the latter ($T_c/J = 0.45$), we show data
for $T/J = 0.04$, 0.08, 0.19 and 0.40 averaged over 16, 22, 26 and 75 samples respectively.

At each temperature there is typically a peak with some breadth, but unlike the two cases with
$x = 0.01$, there is no double-peak structure seen.  This can be understood as increasing
$x$ leading to smaller relative density variations for the Mn spins, and hence smaller
fluctuations in the range of local environments, resulting in narrower distributions
for local charge and magnetization.

\subsection{Correlation of charge and magnetization}
\begin{figure}[htb]
\centerline{\psfig{file=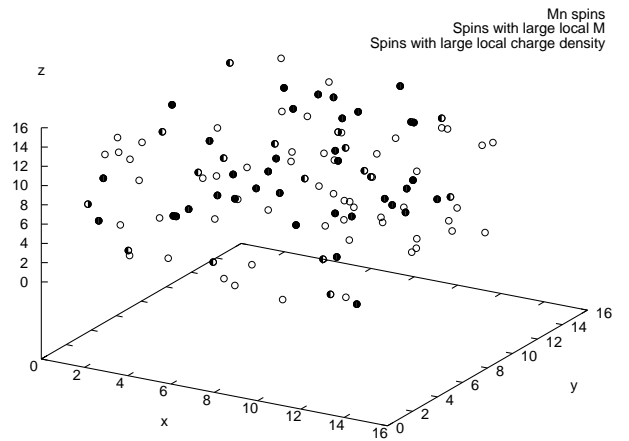,height=7cm,angle=270}}
\caption{Sites with large local magnetization and charge density in a sample with 110 Mn 
spins, 11 carriers at a temperature $T = 0.01 \, J \sim 0.25 \, T_c$.}
\label{3D}
\end{figure}

Large local magnetization of the local Mn spins are correlated with large local charge
densities of the carriers at the Mn site.  To illustrate this point, 
in Fig.~\ref{3D} we show the correlation of large local magnetizations and charge 
densities in a single sample with 110 Mn spins and 11 carriers at a temperature 
$T = 0.01 \, J \simeq 0.25 \, T_c$.  Open circles correspond to sites with low charge density 
$\rho_i < 0.1$ and low magnetization $M^{\rm local}_i < 0.4$.  Solid filled circles correspond
to sites with $\rho_i > 0.1$ and $M^{\rm local}_i > 0.4$  Sites with $\rho_i < 0.1$ but 
$M^{\rm local}_i > 0.4$ or with $\rho_i > 0.1$ and $M^{\rm local}_i < 0.4$ are shown as half-filled
circles.  Less than 14\% of the sites are half-filled indicating a strong correlation between $\rho_i$
and $M^{\rm local}_i$.

\begin{figure}[htb]
\centerline{\psfig{file=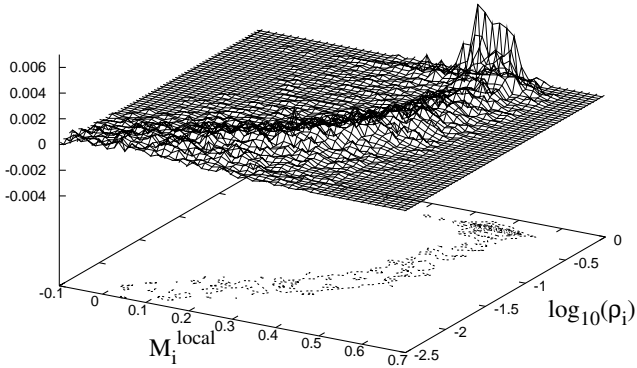,height=7cm,angle=270}}
\caption{Joint distribution function of local magnetization and charge density for $x = 0.01$, $p=0.1$, 
$N_d = 110$, $N_h = 11$
 at temperature $T = 0.01 \, J \simeq T_c/4$.}
\label{jpdf11}
\end{figure}

\begin{figure}[htb]
\centerline{\psfig{file=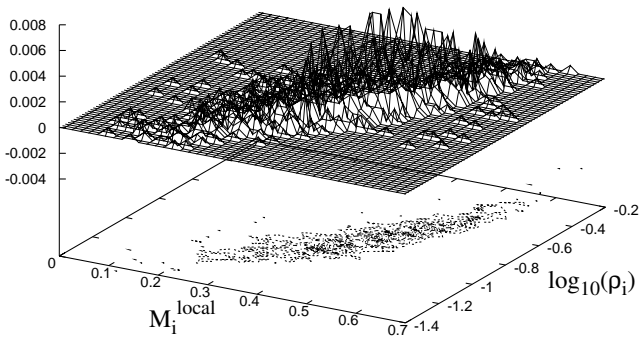,height=7cm,angle=270}}
\caption{Joint distribution function 
of local magnetization and charge density for $x = 0.03$, $p=0.1$, $N_d =41$,
$N_h = 4$ 
 at temperature $T = 0.06 \, J \simeq T_c/3$.}
\label{jpdf31}
\end{figure}

Fig. \ref{jpdf11} shows the joint distribution function for local magnetization and local charge density
for $x = 0.01$, $p = 0.1$.  The data inhabits a narrow band with the peak corresponding to
localized carriers also corresponding to Mn spins with large local magnetizations.
In Fig. \ref{jpdf31} we plot the joint distribution function 
for local magnetization and local charge density
for $x = 0.03$, $p = 0.1$.  The distribution is similar 
to the previous case in that there is also a strong 
correlation between charge density and local magnetization.  However, the data occupy
a smaller region of the 
$M^{\rm local}_i$, $\rho_i$ plane (which is not surprising since both distributions are 
narrower than in the insulating case).

\section{Discussion and Conclusions}
\label{sec:conc}
In this study we have performed a Monte Carlo study of an impurity band model for III-V DMS.
In order to do this we are simulating a model of fermions coupled to classical degrees (spins)
and this requires diagonalization of the fermion problem for every classical
spin configuration.  This is very time-consuming computationally. To 
speed up the procedure, we introduce quantum mechanical perturbation theory coupled with 
Monte Carlo, which we call the PMC scheme. 
The method and tests are described in detail in Sec.~\ref{sec:PMC}. 
Our model restricts charge carriers to 
an impurity band formed from the isolated acceptor impurity states introduced by the Mn ions, 
interacting via antiferromagnetic exchange with Mn spins, which we treat classically.  
This model is based on the picture for the low doping limit of localized carriers. 
Experimental evidence suggests that such an impurity band exists in the vicinity of the metal-insulator
transition (MIT). \cite{Grandidier,ARPES}  The advantage of this type of model in comparison to other 
models that  have been suggested for III-V DMS which start from a valence band point of 
view for the carriers, is that it naturally incorporates the physics associated with the 
MIT, which may be important even in the metallic region.  We have 
considered parameters appropriate for (Ga,Mn)As, although qualitative features of the 
results may well apply for other III-V DMS.

The impurity band model has previously been studied using a mean-field approximation, 
\cite{Mona1,Mona2} in which it was found that the magnetic properties are very unlike those 
of a conventional ferromagnet.  The magnetization curves $S_{Mn}(T)$ were found to be concave 
upwards for a significant portion of the temperatures below $T_c$, 
unlike conventional convex upward curves, and it was predicted that the magnetization
should be inhomogeneous.  The role of disorder in the material was also examined by comparing 
the magnetization curves for ordered arrays of dopants and impurities placed with varying 
degrees of randomness.  It was found that randomly placed impurities led to a higher $T_c$ 
than that found for an ordered lattice of impurities.  However, in the mean field solution, the
electron spin is chosen  to be either parallel or
antiparallel to the total magnetization. Further, mean field approximations neglect temporal 
fluctuations, and are known to overestimate $T_c$.  These overestimations could be significantly
different in the 
ordered and disordered cases.  

The results of the Monte Carlo simulations confirm several of the results of the mean field study,
while differing on some others.  Firstly, the shapes of the magnetization versus temperature curves,
$S_{Mn}(T)$ are found to be unusual, as in the mean field study, but more linear in $T$ compared with
the strongly concave upward curves in the mean field approximation.  Further, the carrier and Mn spin
magnetization follow each other more closely, unlike the mean field, where the carriers remain 
polarized to much higher temperatures than the Mn spins.  We believe this difference is a 
direct consequence of the neglect of temporal fluctuations in the mean field approximation.  The Monte
Carlo data appear
to be much more in accord with experimental results, which suggest that both Mn and carrier spins 
depolarize with
roughly the same temperature dependence.  \cite{Beschoten}  In fact, the magnetization curves
observed in this study bear a striking resemblence to those obtained with magnetic circular 
dichroism results in Ref.~\onlinecite{Beschoten}.  
Whilst magnetic susceptibilities were not considered in the mean field study, in a Monte Carlo study 
applicable to II-VI semiconductors, a peak was found in the ordered phase, well below $T_c$, in 
addition to the singularity at $T_c$. \cite{Xin1,Xin2} Our susceptibility data show a single peak; however, the 
peak is at a temperature significantly below the $T_c$ obtained from the $G(L,T)$ curves.  This is
expected, since our sample sizes are small and the two peaks are not separated as a result.  The 
relative proximity of the low $T$ peak to $T_c$ compared to the II-VI case, makes this more 
difficult to resolve.  Nevertheless,  the explanation of the peak being due to 
free or partially-free spins that are outside the percolating magnetic cluster 
appears to be viable in both cases.

To study the inhomogeneities in the magnetic behaviour identified in the mean field study,
we have studied distributions of local quantities -- charge density and magnetization.
We have considered their joint distribution as well, 
to investigate the correlation between the two.  The calculations of the local charge 
density at low temperatures for the lowest density ($x = 0.01$ and $p = 0.1$), which appears 
to be insulating, are in 
very close  accord with those obtained previously at the mean field level. \cite{Mona2} 
The generic feature that appears to be present is that the local charge density has significant 
dispersion.  For $x = 0.01$ we find a two-peaked structure in the local charge density distribution, 
corresponding to some sites having quasi-localized carriers and others having very low 
charge density, whilst for $x= 0.03$ there is still a broad distribution of charge densities, 
but there appears to be only one  peak, rather than two, indicating that the carriers are in 
general delocalized.  The distributions of local magnetizations appear to reflect the same physical 
picture -- for $x = 0.01$ at temperatures approximately $T_c/2$, there are two peaks, 
which appears to correspond to two  populations of Mn spins -- one which is still 
strongly magnetized, and one which is very slightly
magnetized, whereas for $x = 0.03$ this double-peaked structure is not observed.  The $x = 0.01$
behaviour was previously predicted from mean field calculations, \cite{Mona2,MMR} 
and formed the basis
for a phenomenological model that has been shown to have the capacity to describe experimental 
magnetization curves. \cite{MMR}  Finally, the joint distribution of local charge density and local
magnetization indicate that there is a 
strong correlation between higher charge 
density and large local magnetization. 

One of the main aims of the Monte Carlo study was to discover how important fluctuations are
in determining the critical temperature $T_c$ at various $x$ and $p$.  As shown in Table ~\ref{tab:tc}
mean field theory appears to be more precise in this model for increasing hole concentration, 
which is in contrast with Monte Carlo simulations on a valence band model where the opposite 
trend was observed. \cite{Schliemann,Jungwirth}

\begin{figure}[htb]
\centerline{\psfig{file=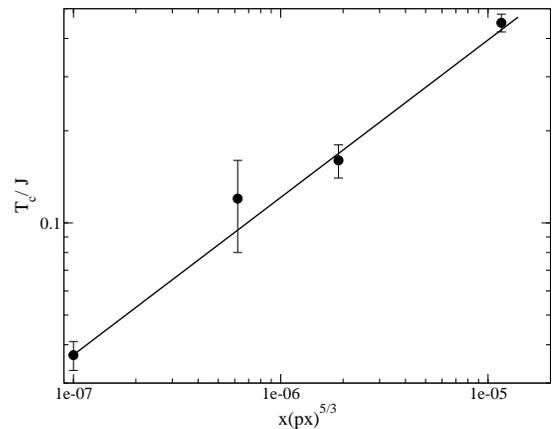,height=6cm,angle=270}}
\caption{$T_c$ as a function of carrier concentration and Mn concentration
 for the four cases considered here.  The line has slope $\frac{1}{2}$}
\label{tcxp}
\end{figure}

A second item of interest is the quantification of the dependence of $T_c$ on the Mn concentration
$x$ and the carrier density $px$.  For individual Mn coupled to free carriers, several models 
exist involving a combination of Fermi energy $E_F \sim (px)^\frac{2}{3}$, and exchange  
energy $E_{ex} \sim Jx$. The generic dependence of $T_c$ is found to be of the form
$T_c \sim x^\alpha (px)^\beta$, where mean field estimates \cite{DOM,Tc,DMFT} give $\alpha = 1$,
with $\beta$ varying between $\frac{1}{3}$ and 1, whilst an analysis involving collective
spin wave excitations with RKKY interactions yields $\alpha = 2$, $\beta = -\frac{1}{3}$ 
for weak coupling ($E_{ex} \ll E_F$) and $\alpha = -\frac{1}{3}$, $\beta =1$ for strong
coupling ($E_{ex} \gg E_F$).  That such a dependence exists for an impurity band is not 
clear; nevertheless in the range of Mn and carrier concentrations studied ($x = 0.01$ - 0.03,
$px =$ 0.001 - 0.009), our $T_c$ can be fit by an expression of the form $T_c \sim x^\alpha
(px)^\beta$ with $\alpha = 0.5 \pm 0.15$, $\beta = 0.85 \pm 0.15$.  As can be seen in Fig. \ref{tcxp}, a double 
logarithmic plot of $T_c/J$ versus $x(px)^\frac{5}{3}$ yields a straight line with a slope of 
$\frac{1}{2}$.  Since the dynamic range in carrier concentration is larger, the data restrict the 
range of allowed $\beta$ more than $\alpha$.  It is interesting that both the dependence on 
carrier density for fixed Mn concentration ($x$), $T_c \sim p^\beta$ and on Mn concentration for 
fixed degree of compensation ($p$), $T_c \sim x^{\alpha + \beta}$, yield exponents 
$\beta = \frac{5}{6}$, $\alpha + \beta = \frac{4}{3}$, which lie within the range predicted
by various treatments for free carriers, for which $\beta$ varies from $-\frac{1}{3}$ to 1 and 
$\alpha + \beta$ from $\frac{2}{3}$ to 2.  Determination of the separate dependences 
on carrier and Mn concentration in experiment would be useful in clarifying the applicability of 
various theoretical models.

Experimental magnetization measurements also 
suggest that there is some degree of variation of local environment for Mn spins, as suggested 
here.  When SQUID magnetometer measurements and anomalous Hall effect measurements are compared,
the magnetization curves that are observed have very different characteristic behaviour.  The 
SQUID measurements tend to be non-conventional magnetization curves as observed here, since they
sample all Mn spins.  However, the Hall effect measurements are made in 
transport and give curves that look quite conventional.  It has been suggested \cite{MMR} that 
this can be understood by the fact that the carriers that are measured in transport preferentially
interact with the Mn spins that are strongly coupled to carriers, so that the magnetic properties 
of only one population of Mn spins are sampled in transport.  The correlation between local
charge density and magnetization seen here supports this scenario.

There are still many experimental questions that remain unresolved about the nature of the 
magnetic behaviour in (Ga,Mn)As.  A more systematic understanding of the dependence of $T_c$ 
on $x$ and $p$ will help determine which models are more appropriate for which regions of the 
phase diagram.  In particular, more accurate determinations of $p$ appear to be one of the most
important ingredients. Recent steps in this direction where the carrier concentration has 
been controlled with electron doping are a start, \cite{holes} but this may introduce other
complications due to dopant centers being spatially distinct from the spins, unlike the case 
for Mn, where the two are in the same place.  Local probes of the material like nuclear 
magnetic resonance will also help to uncover
to what extent the magnetic environment is inhomogeneous, and how this depends on $p$ and $x$.

Despite the qualitative agreements with experimental data, there are a number of effects that 
are left out in the treatment of DMS by our model.\cite{Mona2,Review}  
Within the mean field approximation, a number of hopping 
integrals have been considered within a tight-binding description of the impurity band,\cite{Mona2} 
and it appears that the most important feature in
determining $T_c$ is the density of states at the Fermi energy. \cite{Mona2}  
Other effects that have been left out
in this model are carrier-carrier interactions, valence band states, spin-orbit effects 
and direct Mn-Mn interactions
which could lead to frustration. \cite{frustrate}
Whilst the direct Mn-Mn exchange
should not be important for the $x$ values considered here, since there are relatively 
few Mn spins that are close enough for their antiferromagnetic exchange to be important, 
they may become significantly more important at higher $x \ge 0.1$, and affect $T_c$.

In conclusion, we have performed Monte Carlo simulations on an impurity band model for III-V DMS.
We have confirmed many of the features that were seen at the level of mean field simulations --
unusual magnetization curves and inhomogeneous magnetization and charge density below $T_c$.  The
unusual magnetization curves are also in qualitative agreement with many experiments.  We
have made a comparison of $T_c$ determined using each method and find that for larger values of 
Mn concentration and carrier concentration, the mean field determination of $T_c$ become more 
accurate.  We also find a power-law relation between $T_c$ and carrier concentration and Mn 
concentration, that could be compared with experiments.

%%%%%%%%%%%%%%%%%%%%%%%%%%%%%%%%%%%%%%%%%%%%%%%%%%%%%%%%%%%%%%%%%%%%%%%%%%%%%%

\section{ACKNOWLEDGEMENTS}
We thank Xin Wan for helpful suggestions both in the early stages of this work and during
preparation of the manuscript.  This research was supported by NSF grant DMR-9809483.  M. B. 
acknowledges support from the Natural Sciences and Engineering Research Council of Canada.

\begin{appendix}
\section{Exact solution for the toy model}
\label{app:toy}

Consider the Hamiltonian of Eq.~(\ref{eq:Mona})
\begin{equation}
\label{a1}%%%%%%%%%%%%%%%%%%%%%%%%%%%%%%%%%%%%%%%%
{\cal H}= - t \sum_{\langle i,j\rangle}^{} c_{i\sigma}^{\dagger}
c_{j\sigma} + {J\over N} \bvec{S} \cdot
\sum_{j=1}^{N} c^{\dagger}_{j\alpha} \frac{1}{2} \mbox{\boldmath${\sigma}$}_{\alpha\beta}
 c_{j \beta} ,
\end{equation}
where
$$
\bvec{S} = \sum_{i=1}^{N} \bvec{S}_i ,
$$
is the total spin of the system. 

We parametrize $\bvec{S}=S(\sin{\theta}\cos{\phi},
\sin{\theta}\sin{\phi},\cos{\theta})$ and perform the canonical
transformation
\begin{eqnarray*}
d_{i\uparrow}  & = &\cos{{\theta \over 2}} c_{i\uparrow} + \sin{{\theta
 \over 2}} e^{-i\phi} c_{i\downarrow} ,\\
d_{i\downarrow} & = &- \sin{{\theta \over 2}} e^{i\phi} c_{i\uparrow} +
 \cos{{\theta \over 2}} c_{i\downarrow}.
\end{eqnarray*}
It is then straightforward to show that 
$$
{\cal H} = -t \sum_{\langle i,j \rangle}^{} d_{i\sigma}^{\dagger}
d_{i\sigma} + \frac{JS}{2N} \sum_{i=1}^{N}
\left(d^{\dagger}_{i\uparrow} d_{i\uparrow} - d^{\dagger}_{i\downarrow}
d_{i\downarrow} \right).
$$
As a result, we can diagonalize the Hamiltonian in the $\vec{k}$-space
\begin{equation}
{\cal H}=\sum_{k}^{} E_{k\sigma}(S) d^{\dagger}_{k\sigma} d_{k\sigma} ,
\end{equation}
where
\begin{equation}
E_{k\sigma}(S)= -2t \cos{ka} + { 1 \over 2N} J_oS \sigma ,
\end{equation}
with $\sigma = \pm 1$ and the cyclic boundary conditions 
imply that  $ka = 2\pi n/N$, with $n=0,1,...,N-1$. There is a single lowest 
eigenvalue corresponding to $k=0$, and then degenerate eigenvalues
corresponding to left and right moving modes.

Thus the grand-canonical partition function is [see Eq. (\ref{eq:GCPF})]
\begin{equation}
\label{eq:a2}
Z_N= \prod_{i=1}^{N} \left[\int_{}^{}d \Omega_i \right]\prod_{k\sigma}^{}\left( 1 +
e^{-\beta(E_{k\sigma}(S)-\mu)}\right)  ,
\end{equation}
where we use the simplified notation $\int_{}^{}d \Omega_i =
\int_{0}^{\pi}\sin{\theta_i}d\theta_i \int_{0}^{2\pi}d\phi_i$ for the integral over solid angle. 
One can avoid the $2N$  multiple integrals over
individual spin angles, and replace
them by an expression of the general form
$$
Z_N = \int_{0}^{2\pi} d \phi \int_{0}^{\pi} d\theta \sin{\theta}
\int_{0}^{N} dS S^2 F_N(S,\theta,\phi)
$$
\begin{equation}
\label{eq:a3}
 \times \prod_{k\sigma}^{}( 1 +
e^{-\beta(E_{k\sigma}(S)-\mu)}) , 
\end{equation}
where $\theta,\phi$ define the orientation of the total spin
$\bvec{S}$, and we use the fact that the magnitude of the total spin, $S$, varies between 0
and $N$.
Comparing Eqs. (\ref{eq:a2}) and (\ref{eq:a3}), the definition of the
``weight'' $F_N(\bvec{S})$ is
\begin{equation}
\label{eq:a4}
F_N(\bvec{S})=\prod_{i=1}^{N} \left[\int_{}^{} d\Omega_i\right]
\delta\left(\bvec{S}-\sum_{i=1}^{N} \bvec{S}_i\right),
\end{equation}
from which it is straightforward 
to derive the  recurrence relation
$$
F_N(\bvec{S}) = \int_{}^{}d \Omega_N \, F_{N-1}(\bvec{S}- \bvec{S}_N)
\theta(N-1 - | \bvec{S}- \bvec{S}_N|) ,
$$
where the  Heaviside function $\theta$ insures that the
argument of $F_{N-1}$ cannot  have a magnitude larger than $N-1$. 
It is apparent that, in fact, $F_N(\bvec{S})= F_N(S)$. This can
easily be seen from the definition (\ref{eq:a4}) as well, since one can
choose to define the angles $\Omega_i$ with respect to the system of
coordinates in which $\bvec{S}= S \bvec{\Omega} = S \bvec{z}$,
 and  the  result cannot
depend on the particular orientation ${\theta,\phi}$ of the total spin
$\bvec{S}$. Using the new variable $y = \cos{\theta_N}$, and performing the
trivial integral over $\phi_N$, the recurrence relation can be rewritten as:
$$
F_N(S) = 2 \pi \int_{-1}^{1} dy \, F_{N-1} ( \sqrt{1 +S^2 - 2 S y})
$$
\begin{equation}
\label{eq:a7}
\times \theta( N-1 - \sqrt{1 +S^2 - 2 S y}).
\end{equation}
Defining
\begin{equation}
\label{eq:7b}
F_N(S) = (4\pi)^{N-1} f_N(S),
\end{equation}
 we obtain the recurrence formula
\begin{equation}
\label{eq:a8}
f_N(x) = { 1 \over 2 x} \int_{|x-1|}^{min(x+1, N-1)}dz \, z f_{N-1}(z) .
\end{equation}
This is supplemented by the ``initial'' condition
\begin{equation}
\label{eq:a9}
f_2(x) = { 1 \over 2x} ,
\end{equation}
which can be obtained through direct integration from
Eq. (\ref{eq:a4}). 

From the recurrence relation Eq.~(\ref{eq:a8}), one can determine the general solution for
the weight function $f_N(x)$ for any integer
$N$. This  is a piecewise function, given by
$$
f_N(x) = { 1 \over 2^{N-1} (N-2)! \, x} \sum_{k=0}^{m}(-1)^k C_{N}^{k}
(N-2k-x)^{N-2} 
$$
on the subinterval $N-2(m+1)\le x \le N-2m$ of the support interval
$[0,N]$, with $m$ an integer $0\le m < N/2$ and where $ C_{N}^{k}= N!/(k!(N-k)!)$ is
the appropriate binomial coefficient.

The partition function is thus
$$
Z_N = (4\pi)^N \int_{0}^{N} dS \, S^2 f_N(S)  \prod_{k\sigma}^{}\left( 1 +
e^{-\beta(E_{k\sigma}(S)-\mu)}\right)  ,
$$
where the chemical potential $\mu$ is fixed from the condition for the
average number of fermions:

\begin{eqnarray}
\langle n_{c}\rangle = \int_{0}^{N} dS S^2 f_N(S) \sum_{q\alpha}^{}
n_{q\alpha} {\prod_{k\sigma}^{}\left( 1 + 
e^{-\beta(E_{k\sigma}(S)-\mu)}\right) \over Z_N} , \nonumber 
\end{eqnarray}
where $n_{q\alpha} = [ e^{\beta(E_{q\alpha}(S)-\mu)} +1 ]^{-1}$ are the
occupation numbers of the fermionic levels.  
Since an analytical expression is available for the weight function
$f_N(S)$, these integrals can be evaluated numerically. Once the
chemical potential $\mu$ is known, any other expectation value, such
as the total spin magnitude
$$
\langle S \rangle = { 1 \over Z_N} \int_{0}^{N} dS \, S^{3} f_N(S)
\prod_{k\sigma}^{}\left( 1 + 
e^{-\beta(E_{k\sigma}-\mu)}\right) 
$$
or the total fermionic spin 
$$
\langle |s| \rangle=  \int_{0}^{N} dS S^2
f_N(S) \sum_{q\alpha}^{} {\alpha \over 2}
n_{q\alpha} {\prod_{k\sigma}^{}\left( 1 + 
e^{-\beta(E_{k\sigma}-\mu)}\right) \over Z_N} 
$$
can be computed for any given temperature $T$.
\end{appendix}

%%%%%%%%%%%%%%%%%%%%%%%%%%%%%%%%%%%%%%%%%%%%%%%%%%%%%%%%%%%%%%%%%%%%%%%%%%%%%%
%%%%%%%%%%%%%%%%%%%%%%%%%%%%%%%%%%%%%%%%%%%%%%%%%%%%%%%%%%%%
%
%
%   References.                                
%
%
%%%%%%%%%%%%%%%%%%%%%%%%%%%%%%%%%%%%%%%%%%%%%%%%%%%%%%%%%%%%


\begin{references}
\bibitem{footnote} We find that the $G(L,T)$ curve for $N_d = 40$, $N_h = 4$ for 
$X = 0.01$ and $p =0.1$ does not cross at the same point.  This we believe is partly due
to extremely small system size, and also may suggest that there is an odd-even effect
for small carrier numbers.
\end{references}
\end{document}